\documentclass[aps,prd,reprint,showpacs,amsmath,amssymb,10pt]{revtex4-1}
\usepackage{graphicx}% Include figure files
\usepackage{bm}% bold math
\usepackage{slashed}

\begin{document}
\title{Breit-Wheeler Process in Intense Short Laser Pulses}
\author{K. Krajewska}
\email[E-mail address:\;]{Katarzyna.Krajewska@fuw.edu.pl}
\author{J. Z. Kami\'nski}
\affiliation{Institute of Theoretical Physics, Faculty of Physics, University of Warsaw, Ho\.{z}a 69,
00-681 Warsaw, Poland}
\date{\today}

\begin{abstract}
Energy-angular distributions of electron-positron pair creation in collisions of a laser beam and
a nonlaser photon are calculated using the $S$-matrix formalism. The laser field is modeled as a
finite pulse, similar to the formulation introduced in our recent paper in the context of Compton scattering 
[Phys. Rev. A {\bf 85}, 062102 (2012)]. The nonperturbative regime of pair creation is considered here.
The energy spectra of created particles are compared with the corresponding spectra obtained using
the modulated plane wave approximation for the driving laser field. A very good agreement in these two
cases is observed, provided that the laser pulse is sufficiently long. For short pulse
durations, this agreement breaks down. The sensitivity of pair production to the polarization of 
a driving pulse is also investigated. We show that in the nonperturbative regime, the pair creation yields depend on the 
polarization of the pulse, reaching their maximal values for the linear polarization. Therefore, we focus on 
this case. Specifically, we analyze the dependence of pair creation on the relative configuration of 
linear polarizations of the laser pulse and the nonlaser photon. Lastly, we investigate 
the carrier-envelope phase effect on angular distributions of created particles, suggesting the possibility
of phase control in relation to the pair creation processes.
\end{abstract}

\maketitle

\section{Introduction}
\label{intro}

Due to the recent advances in laser technology, ultrahigh intense laser pulses of short durations 
are routinely produced in contemporary laboratories. This has inspired the fast growth of high field physics
with a renewed interest in studying quantum electrodynamic (QED) processes in powerful laser fields 
(for recent reviews, see, for instance, Refs.~\cite{DiPiazza,Ehlotzky}). A series of laser-based QED experiments 
were performed at the Stanford Linear Accelerator Center (SLAC)~\cite{Bula,Bamber,Kot,Burke}, reporting on nonlinear Compton scattering
and nonlinear electron-positron pair creation processes (with the latter being of particular importance 
in light of this paper). These first experiments have triggered numerous related theoretical studies.
In particular, various proposals for the pair production through the laser-matter~\cite{Mull2,Mull3,Sieczka,DiPia,Kaminski}, 
laser-laser~\cite{DiP,Dunne,Bell,Mocken,Jiang,Grobe}, or laser-photon~\cite{Fofanov,Denisenko,Ivanov,Ivan,Titov2012,Heinzl} 
interactions have been reconsidered, setting the grounds for predictions of avalanchelike QED cascades~\cite{Kirk,Fedotov,Ruhl,Rafelski,Su12}.

\begin{figure}
\includegraphics[width=7cm]{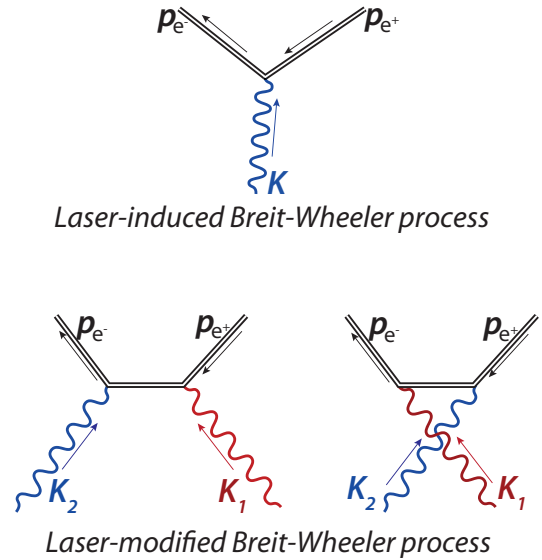}
\caption{(Color online) The Feynman diagrams for the laser-induced (top diagram) and the laser-modified 
(bottom diagrams) Breit-Wheeler processes~\cite{Breit}. The fermion lines (double solid lines) 
represent the Volkov states for electrons and positrons in the background laser field.}
\label{bwqed}
\end{figure}

Generally speaking, strong-field QED processes fall into two categories as they can be either {\it induced} or {\it modified} 
by the laser field. While the laser-induced processes do not occur in the absence of the field, the laser-modified processes
do, even though their properties can change dramatically under the influence of the field. 
By way of illustration, the Feynman diagrams for the laser-induced 
and the laser-modified Breit-Wheeler processes~\cite{Breit}, in which the electron-positron
pairs are created when a laser beam collides with nonlaser photons, are presented in Fig.~\ref{bwqed}. The double solid lines 
are the fermion lines representing the Volkov states~\cite{Volkov} for electrons and positrons in the background laser field.
It is anticipated that for not too intense laser pulses, as compared to the Sauter-Schwinger unit of intensity, the laser-induced process, being 
the lowest order perturbation process, dominates over the laser-modified processes. Exceptions 
can occur for parameters such that the internal fermion line in the bottom diagrams of Fig.~\ref{bwqed} exhibits Oleinik resonances~\cite{Oleinik}. 
A similar situation is met, for instance, in the Bethe-Heitler process represented by the \textit{trident} diagram,
in which pairs are created by virtual photons~\cite{Mull6,KK,KK1,TwoColor,Kasia,Rapid,Ilder} (see, also 
Refs.~\cite{Olei,Roshchupkin,Panek,Rosh}). In the current paper, we study exclusively the laser-induced Breit-Wheeler process.

Originally, the Breit-Wheeler process consisted in a collision of two energetic photons, which combined
their energies to produce the $e^-e^+$ pair~\cite{Breit}. The generalization of this process for the case when
one of the photon sources is a strong laser beam was considered in Refs.~\cite{Reiss,Narozhnyi,Nikishov}.
In these early studies, the laser field was treated as a monochromatic 
plane wave (see, also Refs.~\cite{Ivanov,Ivan,Denisenko}). The process in a weakly nonmonochromatic laser field was investigated in~\cite{Fofanov}
by means of the slowly-varying envelope approximation. The results presented in Ref.~\cite{Fofanov} 
were obtained for an intense laser pulse that includes several oscillations of 
the field. The Breit-Wheeler process induced by a few-cycle pulse but in the perturbative regime
was considered in~\cite{Titov2012}. In both these studies, the driving laser pulse
was circularly polarized, which considerably simplifies the treatment of the problem. 
It is also worth mentioning that the linearly polarized strong laser pulse driving the pair creation was considered in Ref.~\cite{Heinzl},
however with the focus on the mass shift effects.  What is important in the context of the current paper is that
in neither of the aforementioned works was the carrier-envelope phase (CEP) effect on the yields of produced particles analyzed.

In this article, we give a complete description of laser-induced Breit-Wheeler process by finite 
ultrastrong laser pulses, which is along the lines formulated for the Compton scattering in Ref.~\cite{Compton}. 
Our current numerical results are for different pulse durations 
(i.e., for few- and many-cycle laser pulses) in the nonperturbative regime of laser-matter interaction. 
It is expected that in this regime, the yields of produced pairs strongly depend on 
a polarization of the driving laser field, being dominant for the linear polarization. This is the case, 
for instance, for the Bethe-Heitler process~\cite{KK}, and it is also the main focus in this paper.
In our numerical illustrations, we choose the colliding extra photon to be linearly polarized as well. 
Although we refer to it as the $\gamma$-photon, we would like to point out that it has not necessary 
to be the high-energy photon (as it follows from the discussion in Sec.~\ref{invariance}).

This paper is organized as follows. A detailed theoretical description of the laser-induced Breit-Wheeler
process by a single pulse and by an infinite sequence of such pulses is provided in
Secs.~\ref{single} and~\ref{train}, respectively, with the latter being the modulated plane wave field.
We discuss the general kinematic conditions for the Breit-Wheeler process by the finite
pulse in Sec.~\ref{kinematics}, whereas we recapitulate our formulation of this process
(given in Sec.~\ref{single}) to account for the relativistic invariance of QED in Sec.~\ref{invariance}. 
The numerical results presenting the energy-angular spectra of created particles and their
sensitivity to the pulse duration of the driving laser field are studied in Sec.~\ref{compare}. The
validity of the modulated plane wave approximation is also tested in Sec.~\ref{compare}.
The CEP effects in the pair creation spectra are demonstrated in Sec.~\ref{CEP}.
We summarize our results and give prospects for our further investigations in Sec.~\ref{conclusions}.

\section{Theory}
\label{theory}

We use the notation and mathematical convention introduced in our recent publication~\cite{Compton}.
In particular, in formulas we keep $\hbar= 1$ but our numerical results 
are presented in relativistic units such that $\hbar=c = m_{\rm e} = 1$, where $m_{\rm e}$ is the electron
mass. We write $a \cdot b = a^\mu b_\mu (\mu = 0, 1, 2, 3)$ for a product of any two 
four-vectors $a$ and $b$, and $\slashed{a} = \gamma\cdot a = \gamma^\mu a_\mu$ where 
$\gamma^\mu$ are the Dirac gamma matrices. Here, the Einstein summation convention 
is used. We use the so-called light-cone variables defined in Ref.~\cite{Compton}. Namely, for a given space 
direction determined by a unit vector $\bm{n}$ and for an arbitrary four-vector $a$ we keep 
the following notations: $a^\|=\bm{n}\cdot\bm{a}$, $a^-=a^0-a^\|$, $a^+=(a^0+a^\|)/2$, and $\bm{a}^\bot=\bm{a}-a^\|\bm{n}$.
For the four-vectors, we use both the contravariant $(a^0,a^1,a^2,a^3)$ and the standard $(a_0,a_x,a_y,a_z)$ notations.

Using the $S$-matrix formalism, we find that in the lowest order of perturbation theory
(represented in the upper diagram of Fig.~\ref{bwqed}), the probability amplitude for the Breit-Wheeler process, 
$\gamma_{\bm{K}\sigma}\rightarrow e^-_{\bm{p}_{{\rm e}^-}\lambda_{{\rm e}^-}}+e^+_{\bm{p}_{{\rm e}^+}\lambda_{{\rm e}^+}}$, 
with electron and positron momenta and spin polarizations 
$\bm{p}_{{\rm e}^-}\lambda_{{\rm e}^-}$ and $\bm{p}_{{\rm e}^+}\lambda_{{\rm e}^+}$, 
respectively, equals
\begin{multline}
{\cal A}(\gamma_{\bm{K}\sigma}\rightarrow e^-_{\bm{p}_{{\rm e}^-}\lambda_{{\rm e}^-}}+e^+_{\bm{p}_{{\rm e}^+}\lambda_{{\rm e}^+}})=\\
-\mathrm{i}e\int \mathrm{d}^{4}{x}\, j^{(+-)}_{\bm{p}_{{\rm e}^-}\lambda_{{\rm e}^-},
\bm{p}_{{\rm e}^+}\lambda_{{\rm e}^+}}(x)\cdot A^{(+)}_{\bm{K}\sigma}(x), \label{BWAmplitude}
\end{multline}
where $\bm{K}\sigma$ denotes the initial nonlaser photon momentum and polarization. Here,
\begin{equation}
A^{(+)}_{\bm{K}\sigma}(x)=\sqrt{\frac{1}{2\varepsilon_0\omega_{\bm{K}}V}} 
\,\varepsilon_{\bm{K}\sigma}\mathrm{e}^{-\mathrm{i}K\cdot x},
\label{per}
\end{equation}
where $V$ is the quantization volume, $\varepsilon_0$ is the vacuum electric permittivity,
$\omega_{\bm{K}}=cK^0=c|\bm{K}|$ ($K\cdot K=0$), and $\varepsilon_{\bm{K}\sigma}=(0,\bm{\varepsilon}_{\bm{K}\sigma})$ 
is the polarization four-vector satisfying the conditions,
\begin{equation}
K\cdot\varepsilon_{\bm{K}\sigma}=0,\quad \varepsilon_{\bm{K}\sigma}\cdot\varepsilon_{\bm{K}\sigma'}^*=-\delta_{\sigma\sigma'},
\label{polargamma}
\end{equation}
for $\sigma,\sigma'=1,2$. The matrix element of the pair current operator, $j^{(+-)}_{\bm{p}_{{\rm e}^-}\lambda_{{\rm e}^-},\bm{p}_{{\rm e}^+}\lambda_{{\rm e}^+}}(x)$, 
is defined as
\begin{equation}
[j^{(+-)}_{\bm{p}_{{\rm e}^-}\lambda_{{\rm e}^-},\bm{p}_{{\rm e}^+}\lambda_{{\rm e}^+}}(x)]^{\nu}
=\bar{\psi}^{(+)}_{\bm{p}_{{\rm e}^-}\lambda_{{\rm e}^-}}(x)\gamma^\nu \psi^{(-)}_{\bm{p}_{{\rm e}^+}\lambda_{{\rm e}^+}}(x).
\end{equation}
Moreover, $\psi^{(\beta)}_{\bm{p}\lambda}(x)$ (with $\beta=+1$ for electron and $\beta=-1$ for positron) 
is the Volkov solution of the Dirac equation coupled to the electromagnetic field~\cite{Volkov,KK}
\begin{equation}
\psi^{(\beta)}_{\bm{p}\lambda}(x)=\sqrt{\frac{m_{\mathrm{e}}c^2}{VE_{\bm{p}}}}\Bigl(1-\beta\frac{e}{2k\cdot p}\slashed{A}\slashed{k}\Bigr)
u^{(\beta)}_{\bm{p}\lambda}\mathrm{e}^{-\mathrm{i} \beta S_p^{(\beta)}(x)} , \label{Volk}
\end{equation}
with the phase $S_p^{(\beta)}(x)$ having the meaning of the classical action of an electron (positron)
in a laser field,
\begin{equation}
S_p^{(\beta)}(x)=p\cdot x+\int_{-\infty}^{k\cdot x} \Bigl[\beta\frac{ e A(\phi )\cdot p}{k\cdot p}
-\frac{e^2A^{2}(\phi )}{2k\cdot p}\Bigr]{\rm d}\phi .
\label{bbb}
\end{equation}
Here, $E_{\bm{p}}=cp^0\geqslant m_\mathrm{e}c^2$, $p=(p^0,\bm{p})$, $p\cdot p=(m_{\mathrm{e}}c)^2$, and $u^{(\beta)}_{\bm{p}\lambda}$ 
are the free-electron (positron) bispinors normalized such that
\begin{equation}
\bar{u}^{(\beta)}_{\bm{p}\lambda}u^{(\beta')}_{\bm{p}\lambda'}=\beta\delta_{\beta\beta'}\delta_{\lambda\lambda'}.
\end{equation}
The four-vector potential $A(k\cdot x)$ in Eq.~\eqref{Volk} represents an external electromagnetic 
radiation generated by lasers such that $k\cdot A(k\cdot x)=0$ and $k\cdot k=0$. 
Henceforth, we consider the Coulomb gauge for the radiation field which 
means that the four-vector $A(k\cdot x)$ has a vanishing zero component and that the 
electric and magnetic fields are equal to
\begin{align}
\bm{\mathcal{E}}(k\cdot x)&=-\partial_t \bm{A}(k\cdot x)= -ck^0 \bm{A}'(k\cdot x), \label{electric} \\
\bm{\mathcal{B}}(k\cdot x)&=\bm{\nabla}\times \bm{A}(k\cdot x)= -\bm{k}\times \bm{A}'(k\cdot x), \label{magnetic} 
\end{align}
where \textit{'prime'} stands for the derivative with respect to the argument $k\cdot x$. 
In addition, the electric field generated by lasers has to fulfill the following condition
\begin{equation}
\int_{-\infty}^{\infty}\bm{\mathcal{E}}(ck^0t-\bm{k}\cdot\bm{r})\mathrm{d}t=0, \label{LaserCondition1}
\end{equation}
which is equivalent to
\begin{equation}
\lim\limits_{t\rightarrow -\infty}\bm{A}(ck^0t-\bm{k}\cdot\bm{r})=
\lim\limits_{t\rightarrow \infty}\bm{A}(ck^0t-\bm{k}\cdot\bm{r}). \label{LaserCondition2}
\end{equation}
In the next sections, two cases will be considered: when the vector potential $A(k\cdot x)$ describes 
a single laser pulse or an infinite sequence of such pulses; the latter being a modulated
plane wave. Note that each of these situations requires a separate theoretical treatment.

\subsection{Single laser pulse}
\label{single}

In this Section, we formulate theory for the nonlinear Breit-Wheeler process by an isolated laser pulse. 
Assume that a pulse lasts for time $T_\mathrm{p}$, which defines its fundamental frequency 
$\omega=2\pi/T_\mathrm{p}$. The laser field wave four-vector is $k=k^0(1,\bm{n})$, where $k^0=\omega/c$ 
and ${\bm n}$ is a unit vector determining the propagation direction of the laser pulse. 
In accordance with the condition~\eqref{LaserCondition2}, we define the laser pulse such that
\begin{equation}
A(k\cdot x)=0 \quad \textrm{for} \quad k\cdot x < 0\quad \textrm{and}\quad k\cdot x > 2\pi.
\label{LaserVectorPotentialPulse}
\end{equation}
With this definition of the
four-vector potential, it is justified to interpret the momentum $p$ present in the Volkov
solution~\eqref{Volk} as an asymptotic momentum of a free electron (positron).

Keeping this in mind, we consider the most general form of the four-vector potential,
\begin{equation}
A(k\cdot x)=A_0\bigl[\varepsilon_1 f_1(k\cdot x)+\varepsilon_2 f_2(k\cdot x)\bigr], \label{LaserVectorPotential}
\end{equation}
in which two real four-vectors $\varepsilon_i$ describe two linear polarizations and satisfy the relations,
\begin{equation}
\varepsilon_i^2=-1,\quad \varepsilon_1\cdot\varepsilon_2=0,\quad k\cdot\varepsilon_i=0 . \label{polarlaser}
\end{equation}
Note that these conditions do not uniquely determine $\varepsilon_i$. As it can be easily checked, the transformation:
\begin{equation}
\varepsilon_i\rightarrow \varepsilon_i +a_i(k\cdot x)k, \label{gaugek}
\end{equation}
with arbitrary differentiable functions $a_i(k\cdot x)$, does not violate Eqs. \eqref{polarlaser} provided that $k\cdot k=0$, 
which is indeed the case. Note that Eq.~\eqref{gaugek} is a special case of the gauge transformation. The observable quantities, like
the probability distributions that we derive below, should be invariant with respect to such a transformation. We will  
use this fact as a test for our numerical software. The invariance with respect to the gauge 
transformation, Eq. \eqref{gaugek}, permits us to choose $\varepsilon_i$ as the spacelike vectors, since their zeroth-components 
can be nullified by the gauge transformation,
\begin{equation}
\varepsilon_i\rightarrow \varepsilon_i-\frac{\varepsilon_i^0}{k^0}k.
\end{equation}
The same applies to $\varepsilon_{\bm{K}\sigma}$ in Eq. \eqref{polargamma}. For this reason, in our presentation
we choose $\varepsilon_i$ and $\varepsilon_{\bm{K}\sigma}$ as the spacelike four-vectors.

The probability amplitude for the laser-induced Breit-Wheeler process~\eqref{BWAmplitude} can be rewritten as
\begin{equation}
{\cal A}(\gamma_{\bm{K}\sigma}\rightarrow e^-_{\bm{p}_{{\rm e}^-}\lambda_{{\rm e}^-}}+e^+_{\bm{p}_{{\rm e}^+}\lambda_{{\rm e}^+}})
=\mathrm{i}\sqrt{\frac{2\pi\alpha c(m_{\mathrm{e}}c^2)^2}{E_{\bm{p}_{{\rm e}^-}}E_{\bm{p}_{{\rm e}^+}}\omega_{\bm{K}}V^3}}\, \mathcal{A}_{\mathrm{BW}}, \label{ct1}
\end{equation}
where $\alpha=e^2/(4\pi\varepsilon_0 c)$ is the fine-structure constant and
\begin{align}
\mathcal{A}_{\mathrm{BW}}=& \int\mathrm{d}^{4}{x}\mathrm{e}^{-\mathrm{i}(K\cdot x -S^{(+)}_{{p}_{{\rm e}^-}}(x)-S^{(-)}_{{p}_{{\rm e}^+}}(x))} \bar{u}^{(+)}_{\bm{p}_{{\rm e}^-}\lambda_{{\rm e}^-}} \label{ct2} \\
\times  \Bigl(1- & \frac{\mu m_\mathrm{e}c}{2k\cdot p_{{\rm e}^-}}\bigl[f_1(k\cdot x)\slashed{\varepsilon}_1\slashed{k}+f_2(k\cdot x)\slashed{\varepsilon}_2\slashed{k}\bigr]\Bigr) \slashed{\varepsilon}_{\bm{K}\sigma} \nonumber \\
\times  \Bigl(1- & \frac{\mu m_\mathrm{e}c}{2k\cdot p_{{\rm e}^+}} \bigl[f_1(k\cdot x)\slashed{\varepsilon}_1\slashed{k}+f_2(k\cdot x)\slashed{\varepsilon}_2\slashed{k}\bigr]\Bigr) u^{(-)}_{\bm{p}_{{\rm e}^+}\lambda_{{\rm e}^+}}. \nonumber 
\end{align}
Here, we have introduced a relativistically invariant parameter, $\mu=|eA_0|/(m_{\mathrm{e}}c)$,
which measures the intensity of the laser field. After some algebraic manipulations, we find that
a phase present in Eq.~\eqref{ct2} equals
\begin{equation}
K\cdot x-S^{(+)}_{{p}_{{\rm e}^-}}(x)-S^{(-)}_{p_{{\rm e}^+}}(x)
=(K-\bar{p}_{{\rm e}^-}-\bar{p}_{{\rm e}^+})\cdot x+G(k\cdot x),  
\label{ct3}
\end{equation}
with electron and positron laser-field-dressed momenta:
\begin{align}
\bar{p}_{{\rm e}^\mp}=p_{{\rm e}^\mp} \mp & \mu m_\mathrm{e} c\Bigl(\frac{\varepsilon_1\cdot p_{{\rm e}^\mp}}{k\cdot p_{{\rm e}^\mp}}\langle f_1\rangle 
+ \frac{\varepsilon_2\cdot p_{{\rm e}^\mp}}{k\cdot p_{{\rm e}^\mp}}\langle f_2\rangle\Bigr)k \nonumber \\ 
+ & \frac{1}{2}(\mu m_\mathrm{e} c)^2\frac{\langle f_1^2\rangle+\langle f_2^2\rangle}{k\cdot p_{{\rm e}^\mp}}k .  \label{cp1}
\end{align}
Here, we understand that
\begin{equation}
\langle f_i^j \rangle=\frac{1}{T_{\mathrm{p}}}\int\limits_0^{T_{\mathrm{p}}}\mathrm{d}t 
[f_i(ck^0t-\bm{k}\cdot\bm{r})]^j=\frac{1}{2\pi}\int\limits_0^{2\pi}\mathrm{d}\phi [f_i(\phi)]^j, 
\label{ct6}
\end{equation}
for $i, j=1,2$. In Eq.~\eqref{ct3}, we have also introduced,
\begin{align}
G(k\cdot x)&=\int_0^{k\cdot x}\mathrm{d}{\phi}\Bigl[ -\mu m_\mathrm{e} c\Bigl(\frac{\varepsilon_1\cdot p_{{\rm e}^+}}{k\cdot p_{{\rm e}^+}} - \frac{\varepsilon_1\cdot p_{{\rm e}^-}}{k\cdot p_{{\rm e}^-}}\Bigr)\nonumber\\
&\times \bigl(f_1(\phi) -\langle f_1\rangle\bigr) -\mu m_\mathrm{e} c\Bigl(\frac{\varepsilon_2\cdot p_{{\rm e}^+}}{k\cdot p_{{\rm e}^+}} - \frac{\varepsilon_2\cdot p_{{\rm e}^-}}{k\cdot p_{{\rm e}^-}}\Bigr)\nonumber\\
&\times\bigl(f_2(\phi)-\langle f_2\rangle\bigr) 
 -\frac{1}{2}(\mu m_\mathrm{e} c)^2\Bigl(\frac{1}{k\cdot p_{{\rm e}^+}} + \frac{1}{k\cdot p_{{\rm e}^-}}\Bigr)\nonumber\\
&\times\bigl(f_1^2(\phi)-\langle f_1^2\rangle + f_2^2(\phi)-\langle f_2^2\rangle\bigr) \Bigr],  \label{cp4}
\end{align}
which is a periodic function of the argument $k\cdot x$, meaning that $G(0)=G(2\pi)=0$. Using the Fourier expansion,
\begin{equation}
[f_1(k\cdot x)]^n[f_2(k\cdot x)]^m \mathrm{e}^{-\mathrm{i}G(k\cdot x)}
=\sum_{N=-\infty}^\infty G^{(n,m)}_N \mathrm{e}^{-\mathrm{i}Nk\cdot x}, \label{ct7}
\end{equation}
we can perform the space-time integration in Eq. \eqref{ct2}, provided that $n$ and $m$ are not simultaneously equal to 0.
As we have already argued in Ref.~\cite{Compton} when studying the Compton scattering, the case when $n=m=0$
requires a separate treatment. This becomes clear when the light-cone coordinates are used.
The point being that, only if $n\neq 0$ or $m\neq 0$, the integral with respect to $x^-$ becomes limited to the finite region, 
$0\leqslant x^-\leqslant 2\pi/k^0$. In order to treat the case when both $n$ and $m$ are 0, we use the Boca-Florescu
transformation (see, Ref.~\cite{Compton,Boca}) which states that
\begin{multline}
\int\mathrm{d}^{4}{x}\,\exp\Bigl(-\mathrm{i}Q\cdot x-\mathrm{i}\int_{-\infty}^{k\cdot x}\mathrm{d}\phi\,h(\phi)\Bigr) \\
=-\frac{k^0}{Q^0}\int\mathrm{d}^{4}{x} 
\,h(k\cdot x)\exp\Bigl(-\mathrm{i}Q\cdot x-\mathrm{i}\int_{-\infty}^{k\cdot x}\mathrm{d}\phi\,h(\phi)\Bigr) ,  
\label{BocaFlorescu}
\end{multline}
if $Q_0\neq 0$. In our case, 
\begin{eqnarray}
Q&=&K-p_{{\rm e}^-}-p_{{\rm e}^+},\\
h(\phi)&=&a_1 f_1(\phi)+ a_2 f_2(\phi)+b[f^2_1(\phi)+f^2_2(\phi)] ,
\end{eqnarray}
with
\begin{align}
a_1=& -\mu m_{\mathrm{e}}c\Bigl(\frac{\varepsilon_1\cdot p_{{\rm e}^+}}{k\cdot p_{{\rm e}^+}} - \frac{\varepsilon_1\cdot p_{{\rm e}^-}}{k\cdot p_{{\rm e}^-}}\Bigr)=-Q^0\tilde{a}_1/k^0 , \nonumber \\
a_2=& -\mu m_{\mathrm{e}}c\Bigl(\frac{\varepsilon_2\cdot p_{{\rm e}^+}}{k\cdot p_{{\rm e}^+}} - \frac{\varepsilon_2\cdot p_{{\rm e}^-}}{k\cdot p_{{\rm e}^-}}\Bigr)=-Q^0\tilde{a}_2/k^0 , \nonumber \\
b=& -\frac{1}{2}(\mu m_{\mathrm{e}}c)^2 \Bigl(\frac{1}{k\cdot p_{{\rm e}^+}} + \frac{1}{k\cdot p_{{\rm e}^-}}\Bigr)=-Q^0\tilde{b}/k^0 .
\label{tilded}
\end{align}
Using the Fourier decomposition~\eqref{ct7} and applying the Boca-Florescu transformation~\eqref{BocaFlorescu},
we arrive at
\begin{equation}
\mathcal{A}_{\mathrm{BW}}=\sum_N D_N \int\mathrm{d}^{4}{x} \mathrm{e}^{-\mathrm{i}(K-\bar{p}_{{\rm e}^-}-\bar{p}_{{\rm e}^+}+Nk)\cdot x}, \label{ct8}
\end{equation}
where
\begin{align}
D_N=&\bar{u}^{(+)}_{\bm{p}_{{\rm e}^-}\lambda_{{\rm e}^-}}\slashed{\varepsilon}_{\bm{K}\sigma} u^{(-)}_{\bm{p}_{{\rm e}^+}\lambda_{{\rm e}^+}} G^{(0,0)}_N \label{ct9}  \\
 - & \frac{1}{2}\mu m_\mathrm{e} c\Bigl[\Bigl( \frac{1}{k\cdot p_{{\rm e}^+}} \bar{u}^{(+)}_{\bm{p}_{{\rm e}^-}\lambda_{{\rm e}^-}}\slashed{\varepsilon}_{\bm{K}\sigma}  \slashed{\varepsilon}_1\slashed{k} u^{(-)}_{\bm{p}_{{\rm e}^+}\lambda_{{\rm e}^+}}
 \nonumber \\ &\qquad
 +\frac{1}{k\cdot p_{{\rm e}^-}} \bar{u}^{(+)}_{\bm{p}_{{\rm e}^-}\lambda_{{\rm e}^-}} \slashed{\varepsilon}_1\slashed{k}\slashed{\varepsilon}_{\bm{K}\sigma}  u^{(-)}_{\bm{p}_{{\rm e}^+}\lambda_{{\rm e}^+}}\Bigr)G^{(1,0)}_N  \nonumber \\
 &\qquad +\Bigl( \frac{1}{k\cdot p_{{\rm e}^+}} \bar{u}^{(+)}_{\bm{p}_{{\rm e}^-}\lambda_{{\rm e}^-}} \slashed{\varepsilon}_{\bm{K}\sigma} \slashed{\varepsilon}_2\slashed{k} u^{(-)}_{\bm{p}_{{\rm e}^+}\lambda_{{\rm e}^+}}  \nonumber \\ &\qquad
+\frac{1}{k\cdot p_{{\rm e}^-}} \bar{u}^{(+)}_{\bm{p}_{{\rm e}^-}\lambda_{{\rm e}^-}} \slashed{\varepsilon}_2\slashed{k}\slashed{\varepsilon}_{\bm{K}\sigma}  u^{(-)}_{\bm{p}_{{\rm e}^+}\lambda_{{\rm e}^+}}\Bigr)G^{(0,1)}_N\Bigr]  \nonumber \\
 + & \frac{(\mu m_\mathrm{e} c)^2}{4(k\cdot p_{{\rm e}^+})(k\cdot p_{{\rm e}^-})} \Bigl[\bar{u}^{(+)}_{\bm{p}_{{\rm e}^-}\lambda_{{\rm e}^-}} \slashed{\varepsilon}_1\slashed{k} \slashed{\varepsilon}_{\bm{K}\sigma} \slashed{\varepsilon}_1\slashed{k} u^{(-)}_{\bm{p}_{{\rm e}^+}\lambda_{{\rm e}^+}}G^{(2,0)}_N  
 \nonumber \\ &\qquad
+\bar{u}^{(+)}_{\bm{p}_{{\rm e}^-}\lambda_{{\rm e}^-}} \slashed{\varepsilon}_2\slashed{k} \slashed{\varepsilon}_{\bm{K}\sigma} \slashed{\varepsilon}_2\slashed{k} u^{(-)}_{\bm{p}_{{\rm e}^+}\lambda_{{\rm e}^+}}G^{(0,2)}_N  \nonumber \\
 & \qquad +\Bigl(\bar{u}^{(+)}_{\bm{p}_{{\rm e}^-}\lambda_{{\rm e}^-}} \slashed{\varepsilon}_1\slashed{k} \slashed{\varepsilon}_{\bm{K}\sigma} \slashed{\varepsilon}_2\slashed{k} u^{(-)}_{\bm{p}_{{\rm e}^+}\lambda_{{\rm e}^+}}
 \nonumber \\ &\qquad
  +\bar{u}^{(+)}_{\bm{p}_{{\rm e}^-}\lambda_{{\rm e}^-}} \slashed{\varepsilon}_2\slashed{k} \slashed{\varepsilon}_{\bm{K}\sigma} \slashed{\varepsilon}_1\slashed{k} u^{(-)}_{\bm{p}_{{\rm e}^+}\lambda_{{\rm e}^+}}  \Bigr)G^{(1,1)}_N \Bigr] \nonumber
\end{align}
and
\begin{equation}
G^{(0,0)}_N \rightarrow\tilde{a}_1 G^{(1,0)}_N+ \tilde{a}_2 G^{(0,1)}_N+\tilde{b}[G^{(2,0)}_N+G^{(0,2)}_N],
\label{subst}
\end{equation}
with $\tilde{a}_1,\tilde{a}_2,$ and $\tilde{b}$ defined by Eqs.~\eqref{tilded}.

Performing the space-time integration in Eq. \eqref{ct8} and keeping in mind 
that $0\leqslant x^- \leqslant 2\pi/k^0$, we obtain
\begin{equation}
\mathcal{A}_{\mathrm{BW}}=\sum_N (2\pi)^3\delta^{(1)}(P_N^-)\delta^{(2)}(\bm{P}_N^\bot)D_N\frac{1-\mathrm{e}^{-2\pi\mathrm{i}P_N^+/k^0}}{\mathrm{i}P_N^+} ,  \label{cp5}
\end{equation}
where we define
\begin{equation}
P_N=K-\bar{p}_{{\rm e}^-}-\bar{p}_{{\rm e}^+}+Nk. \label{cp6}
\end{equation}
To solve the momentum conservation conditions imposed by the delta functions in~\eqref{cp5}, we assume
that the final positron momentum is known. This allows us to introduce the four-vector $w=K-p_{{\rm e}^+}$, so that
\begin{equation}
p_{{\rm e}^-}^0=p_{{\rm e}^-}^{\|}+w^-, \quad \bm{p}_{{\rm e}^-}^{\bot}=\bm{w}^{\bot}. \label{cp7}
\end{equation}
Since the electron mass is different from zero, it follows from the first of these equations that $w^->0$, and
\begin{equation}
p_{{\rm e}^-}^{\|}=\frac{(m_{\mathrm{e}}c)^2+(\bm{w}_{\bot})^2-(w^-)^2}{2w^-} =\frac{K\cdot p_{{\rm e}^+}}{w^-}+w^{\|} , \label{cp8}
\end{equation}
which means that
\begin{equation}
Q^0=K^0-p_{{\rm e}^-}^0-p_{{\rm e}^+}^0=K^{\|}-p_{{\rm e}^-}^{\|}-p_{{\rm e}^+}^{\|}=-\frac{K\cdot p_{{\rm e}^+}}{w^-}<0 . \label{cp9}
\end{equation}
This is the applicability condition for the Boca-Florescu transformation~\cite{Compton,Boca}.
It also shows that this transformation, Eq. \eqref{BocaFlorescu}, is relativistically invariant, as the ratio
\begin{equation}
\frac{Q^0}{k^0}=-\frac{K\cdot p_{\mathrm{e}^+}}{k\cdot w} =-\frac{K\cdot p_{\mathrm{e}^+}}{k\cdot p_{\mathrm{e}^-}}
=-\frac{K\cdot p_{\mathrm{e}^-}}{k\cdot p_{\mathrm{e}^+}}  \label{ratioinv}
\end{equation}
does not depend on the reference frame.

Note that $\bm{P}_N^\bot$ and $P_N^{-}$ do not depend explicitly on $N$, and
\begin{equation}
\int\mathrm{d}^{3}p_{{\rm e}^-}\,\delta^{(1)}(P_N^{-})\delta^{(2)}(\bm{P}_N^\bot)=\frac{k^0p_{{\rm e}^-}^0}{k\cdot p_{{\rm e}^-}} .
\end{equation}
Taking this into account, we derive that the differential probability distribution of created
positrons in the Breit-Wheeler process induced by a finite laser pulse is
\begin{align}
\frac{\mathrm{d}^{3}{\mathsf{P}^{(\mathrm{p})}}}{\mathrm{d}{E_{\bm{p}_{{\rm e}^+}}}
\mathrm{d}^{2}{\Omega_{\bm{p}_{{\rm e}^+}}}} = &\sum_{\lambda_{{\rm e}^-},\lambda_{{\rm e}^+}=\pm}|\bm{p}_{{\rm e}^+}|\frac{\alpha (m_{\mathrm{e}}c)^2 k^0}{(2\pi)^2 \omega_{\bm{K}}
(k\cdot p_{{\rm e}^-})} \label{cp10} \\  \times & \Big|\sum_N D_N\frac{1-\mathrm{e}^{-2\pi\mathrm{i}P_N^0/k^0}}{P_N^0} \Big|^2 , \nonumber
\end{align}
with the electron four-momentum $p_{{\rm e}^-}$ equal to
\begin{align}
\bm{p}_{{\rm e}^-}^{\bot} &=\bm{w}^{\bot} , \\
p_{{\rm e}^-}^{\|} &= \frac{(m_{\mathrm{e}}c)^2+(\bm{w}^{\bot})^2-(w^-)^2}{2w^-} ,  \\
p_{{\rm e}^-}^0    &= \frac{(m_{\mathrm{e}}c)^2+(\bm{w}^{\bot})^2+(w^-)^2}{2w^-} . 
\end{align}
One can check that the above solution of the momentum conservation conditions, $P_N^-=0$ and ${\bm P}_N^\perp={\bm 0}$,
satisfies the on-mass shell relation, $p_{{\rm e}^-}\cdot p_{{\rm e}^-}=(m_{\mathrm{e}}c)^2$.

If an incident laser pulse is long enough, terms such that $P_N^+\simeq0$ contribute the most in Eq.~\eqref{cp5}.
Hence, it is possible to define the effective energy 
$N_{\mathrm{eff}}ck^0$ absorbed from the laser pulse such that $P_{N_{\mathrm{eff}}}^0=0$.
In this case, 
\begin{equation}
N_{\mathrm{eff}}=\frac{\bar{p}_{\mathrm{e}^-}^0+\bar{p}_{\mathrm{e}^+}^0-K^0}{k^0} =cT_{\mathrm{p}}\frac{\bar{p}_{\mathrm{e}^-}^0+\bar{p}_{\mathrm{e}^+}^0-K^0}{2\pi} . \label{cp12}
\end{equation}
This allows us to define the differential probability of pair creation,
\begin{equation}
\frac{\mathrm{d}^{3}\mathsf{P}^{(\mathrm{p})}}{\mathrm{d}{N_{\mathrm{eff}}}\mathrm{d}^{2}{\Omega_{\bm{p}_{{\rm e}^+}}}}=
\frac{\mathrm{d}E_{\bm{p}_{{\rm e}^+}}}{\mathrm{d}N_\mathrm{eff}} \frac{\mathrm{d}^{3}\mathsf{P}^{(\mathrm{p})}}{\mathrm{d}{E_{\bm{p}_{{\rm e}^+}}}\mathrm{d}^{2}{\Omega_{\bm{p}_{{\rm e}^+}}}} , 
\label{bwpulse} 
\end{equation}
which will have its analog with the case when a train of pulses, instead of a single pulse,
is used. In a very similar way we derive the energy-angular distribution for created electrons, 
the explicit form of which is not presented here.

Note that the laser-dressed momenta~\eqref{cp1} 
are gauge-dependent, as they change their values if the gauge transformation, Eq. \eqref{gaugek}, 
with a constant $a_i$ is applied. For the first time, such a definition of the laser-dressed
momenta has been introduced in the context of Compton scattering in Ref.~\cite{Compton}. 
A similar definition of the laser-dressed momenta 
has been used recently by Harvey {\it et al.}~\cite{Ilderton} for the classical analogue of this process, i.e., 
Thompson scattering. At the same time, we would like to note that in the formulation presented above
only the sum $\bar{p}_{\textrm{e}^-}+\bar{p}_{\textrm{e}^+}$ appears. The point being that this sum 
and, hence, also $N_{\mathrm{eff}}$ [Eq.~\eqref{cp12}] are gauge-invariant. We can redefine the laser-dressed
momenta such that
\begin{equation}
\bar{p}_{\mathrm{e}^{\mp}}\rightarrow \bar{p}_{\mathrm{e}^{\mp}} \mp s_0 k \mp s_1
\varepsilon_1 \mp s_2 \varepsilon_2 , \label{pdressedgaugeinv} 
\end{equation}
with, in principle, arbitrary real constants $s_i$ ($i=0,1,2$). These constants can still depend 
on different kinds of time-averaged shape functions, like $\langle f_1\rangle$, $\langle f_2\rangle$, $\langle f_1f_2\rangle$, etc., 
however they should vanish for vanishing laser field in order to keep the correspondence to the free particle case. 
Note that the new definition~\eqref{pdressedgaugeinv} preserves the sum $\bar{p}_{\textrm{e}^-}+\bar{p}_{\textrm{e}^+}$. 
Therefore, one can use this arbitrariness to define the gauge-invariant laser-dressed momenta. 
By making the substitutions $\varepsilon_i\rightarrow\varepsilon_i+a_i k$, we find that the dressed 
momenta do not depend on $a_i$ provided that $s_1=-\mu m_{\mathrm{e}}c\langle f_1\rangle$ and $s_2=-\mu m_{\mathrm{e}}c\langle f_2\rangle$, with still an arbitrary $s_0$. 
Let us stress, however, that such modifications of the momentum dressing do not affect any observable quantity,
and should be considered only as a mathematical operation. In closing this Section,
let us also emphasize that $N_{\mathrm{eff}}$ [Eq.~\eqref{cp12}] is not only gauge-invariant but it is also relativistic-invariant, since
\begin{gather}
N_{\mathrm{eff}}=-\frac{Q^0}{k^0} + \mu m_{\mathrm{e}}c \bigl( \langle f_1\rangle \varepsilon_1+\langle f_2\rangle \varepsilon_2 \bigr)
\Bigl(\frac{p_{\mathrm{e}^+}}{k\cdot p_{\mathrm{e}^+}}-\frac{p_{\mathrm{e}^-}}{k\cdot p_{\mathrm{e}^-}}\Bigr)
\nonumber \\
 + \frac{1}{2}(\mu m_{\mathrm{e}}c)^2\bigl(\langle f_1^2\rangle +\langle f_2^2\rangle \bigr)
 \Bigl(\frac{1}{k\cdot p_{\mathrm{e}^+}}+\frac{1}{k\cdot p_{\mathrm{e}^-}} \Bigr),
\end{gather}
where $Q^0/k^0$ is defined by Eq. \eqref{ratioinv}.

\subsection{Train of laser pulses}
\label{train}

In this Section, we consider the Breit-Wheeler process by a 
train of laser pulses. We assume that the duration of a single pulse within 
such a field is $T_{\mathrm{p}}$. This defines the fundamental frequency of the
laser field, $\omega=2\pi/T_{\mathrm{p}}$, and also the four-vector
$k=k^0(1,\bm{n})$ with $k^0=\omega/c$ and the propagation direction vector ${\bm n}$.
We consider a laser field described by the four-vector potential
\eqref{LaserVectorPotential}, with the polarization four-vectors as before. 
This time the shape functions $f_i(\phi)$ ($i=1,2$) are periodic functions of their 
argument $\phi$, with the period of $2\pi$. This means that $\langle f_i\rangle=0$,
and the laser-dressed four-momenta of created particles read now
\begin{equation}
\bar{p}_{{\rm e}^\mp}=p_{{\rm e}^\mp} +\frac{1}{2}(\mu m_\mathrm{e} c)^2
\frac{\langle f_1^2\rangle+\langle f_2^2\rangle}{k\cdot p_{{\rm e}^\mp}}k, \label{ct4}
\end{equation}
again up to terms $\mp s_0k$ [see, Eq.~\eqref{pdressedgaugeinv} and the following discussion].

When calculating the probability rate for the nonlinear Breit-Wheeler process by
a laser pulse train, we proceed along the same lines as in Sec.~\ref{single}.
This leads us to expression~\eqref{ct8}, with Eq.~\eqref{ct9} defining the Fourier
components $D_N$ where the substitution~\eqref{subst} no longer applies.
Performing the space-time integration in Eq. \eqref{ct8}, we end up with 
\begin{equation}
 \mathcal{A}_{\mathrm{BW}}=(2\pi)^4\sum_N D_N \delta^{(4)}(K-\bar{p}_{{\rm e}^-}-\bar{p}_{{\rm e}^+}+Nk),
\end{equation}
where the $\delta$-function determines the four-momenta conservation condition.
Hence, the probability of pair creation by a single pulse from the train 
can be calculated,
\begin{align}
\mathsf{P}^{(\mathrm{t})}&=\frac{\alpha T_{\mathrm{p}}}{2\pi}\sum_{\lambda_{{\rm e}^-},\lambda_{{\rm e}^+}=\pm} 
\sum_{N=-\infty}^{\infty}\int \mathrm{d}^3p_{{\rm e}^-}\mathrm{d}^3p_{{\rm e}^+} \label{rownanie} \\ 
&\times\frac{(m_\mathrm{e}c^2)^2c^2}{E_{\bm{p}_{{\rm e}^-}}E_{\bm{p}_{{\rm e}^+}}\omega_{\bm{K}}}|D_N|^2\delta^{(4)}(K-\bar{p}_{{\rm e}^-}-\bar{p}_{{\rm e}^+}+Nk).\nonumber
\end{align}
Due to the conservation condition, four integrations can be carried out in Eq.~\eqref{rownanie}. Let us assume that we know 
the ejection angles of positrons. This means that we can define the angular distribution of positrons
created in the process accompanied by the absorption ($N>0$) or emission ($N<0$) of the laser radiation 
energy $Nck^0$,
\begin{align}
\frac{\mathrm{d}^2\mathsf{P}^{(\mathrm{t})}_N}{\mathrm{d}\Omega_{\bm{p}_{{\rm e}^+}}}&=\frac{\alpha T_{\mathrm{p}}}{2\pi}
\sum_{\lambda_{{\rm e}^-},\lambda_{{\rm e}^+}=\pm}\int \mathrm{d}E_{\bm{p}_{{\rm e}^+}}\mathrm{d}^3p_{{\rm e}^-}  \\ 
&\times |\bm{p}_{{\rm e}^+}|\frac{(m_\mathrm{e}c^2)^2}{E_{\bm{p}_{{\rm e}^-}}\omega_{\bm{K}}}|D_N|^2 \delta^{(4)}(K-\bar{p}_{{\rm e}^-}-\bar{p}_{{\rm e}^+}+Nk). \nonumber
\end{align}
The integration over the momentum $\bm{p}_{{\rm e}^-}$ leads to
\begin{equation}
\int \mathrm{d}^3p_{{\rm e}^-} \delta^{(3)}({\bm K}-\bar{{\bm p}}_{{\rm e}^-}-\bar{{\bm p}}_{{\rm e}^+}+N{\bm k})
= \Big |\frac{\partial\bm{p}_{{\rm e}^-}}{\partial\bar{\bm{p}}_{{\rm e}^-}}\Big |=\frac{p_{{\rm e}^-}^0}{\bar{p}_{{\rm e}^-}^0}.
\end{equation}
For the remaining integral, we obtain
\begin{equation}
\int \mathrm{d}E_{\bm{p}_{{\rm e}^+}}\delta^{(1)}(K^0-\bar{p}_{{\rm e}^-}^0-\bar{p}_{{\rm e}^+}^0+Nk^0)=\sum_{\ell}\frac{c}{D^{(\ell)}(\hat{\bm{p}}_{{\rm e}^+})}
\end{equation} 
where
\begin{eqnarray}
D^{(\ell)}(\hat{\bm{p}}_{{\rm e}^+}) &=& \biggl|-1+\frac{p_{{\rm e}^+}^0}{\bar{p}_{{\rm e}^-}^0}\frac{{\bm p}_{{\rm e}^+}\cdot\bar{\bm p}_{{\rm e}^-}}{|{\bm p}_{{\rm e}^+}|^2}+(\bar{p}_{{\rm e}^+}^0-p_{{\rm e}^+}^0)\nonumber\\
&\times&\frac{k\cdot{p}_{{\rm e}^-}}{\bar{p}_{{\rm e}^-}^0(k\cdot p_{{\rm e}^+})}\biggl(1-\frac{p_{{\rm e}^+}^0}{k^0}\frac{{\bm k}\cdot{\bm p}_{{\rm e}^+}}{|{\bm p}_{{\rm e}^+}|^2}\biggr)\biggr|,
\end{eqnarray} 
and $\ell$ labels all possible solutions of the conservation condition~\cite{KK}. 
Finally,
\begin{equation}
\frac{\mathrm{d}^2\mathsf{P}^{(\mathrm{t})}_N}{\mathrm{d}\Omega_{\bm{p}_{{\rm e}^+}}}=
\alpha\frac{(m_{\rm e}c)^2}{k^0K^0}\sum_{\lambda_{{\rm e}^-},\lambda_{{\rm e}^+}=\pm}\sum_{\ell} \frac{|\bm{p}_{{\rm e}^+}|}{\bar{p}_{{\rm e}^-}^0
D^{(\ell)}(\hat{\bm{p}}_{{\rm e}^+})}|D_N|^2 .
\end{equation}
In order to be able to compare this result with the case of a single laser pulse
discussed in Sec.~\ref{single}, let us define symbolically the triple differential 
probability distribution of pair creation,
\begin{equation}
\frac{\mathrm{d}^3\mathsf{P}^{(\mathrm{t})}}{\mathrm{d}N\mathrm{d}\Omega_{\bm{p}_{{\rm e}^+}}}
=\frac{\mathrm{d}^2\mathsf{P}^{(\mathrm{t})}_N}{\mathrm{d}\Omega_{\bm{p}_{{\rm e}^+}}}. \label{bwtrain}
\end{equation}
This quantity corresponds to Eq.~\eqref{bwpulse} defined in Sec.~\ref{single}.

\subsection{Remarks on kinematics}
\label{kinematics}

Going back to Sec.~\ref{single}, where the $e^-e^+$ pair creation induced by a finite laser pulse
was considered, we note that the condition $w^->0$ imposes some constraints on possible values of the positron 
final momenta and the geometry of the process. Let us choose the system of coordinates such that the $z$-axis 
is determined by the propagation direction of the laser field, $\bm{n}=\bm{e}_z$. Therefore, the 
inequality $w^->0$ can be put down as
\begin{equation}
|\bm{p}_{\mathrm{e}^+}|^2\sin^2\theta_{\mathrm{e}^+}-2K^-|\bm{p}_{\mathrm{e}^+}|\cos\theta_{\mathrm{e}^+} +(m_{\mathrm{e}}c)^2-(K^-)^2<0 ,
\end{equation} 
where $\theta_{\mathrm{e}^+}$ is the positron polar angle measured with respect to the $z$-axis.
This inequality has real solutions for $|\bm{p}_{\mathrm{e}^+}|$ provided that
\begin{equation}
\sin\theta_{\mathrm{e}^+}<\frac{\sqrt{(K^-)^2}}{m_{\mathrm{e}}c}. \label{kinlimitation1}
\end{equation}
We conclude from here that $K^->0$, which excludes the situation when the incident photon 
and the laser field propagate in the same direction.
Defining
\begin{align}
p_{\mathrm{min}}&=\frac{K^-\cos\theta_{\mathrm{e}^+}-\sqrt{(K^-)^2- (m_{\mathrm{e}}c)^2\sin^2\theta_{\mathrm{e}^+}}}{\sin^2\theta_{\mathrm{e}^+}} , \\
p_{\mathrm{max}}&=\frac{K^-\cos\theta_{\mathrm{e}^+}+\sqrt{(K^-)^2- (m_{\mathrm{e}}c)^2\sin^2\theta_{\mathrm{e}^+}}}{\sin^2\theta_{\mathrm{e}^+}} ,
\end{align}
we find that
\begin{equation}
\max (0,p_{\mathrm{min}}) < |\bm{p}_{\mathrm{e}^+}| < p_{\mathrm{max}} ,
\end{equation}
provided that $p_{\mathrm{max}}>0$; otherwise, the $\gamma$-photon cannot create the $e^-e^+$ pair. 
In particular, if $\cos\theta_{\mathrm{e}^+}\geqslant 0$ the pair can always be created [under the 
circumstances that the inequality~\eqref{kinlimitation1} still remains valid], whereas for $\cos\theta_{\mathrm{e}^+} < 0$ 
the pair can be formed only for a sufficiently energetic nonlaser photon,
\begin{equation}
\omega_{\bm{K}}>\frac{m_{\mathrm{e}}c^2}{1-\bm{n}\cdot\bm{n}_{\bm{K}}} .
\end{equation}
It follows from this condition that the least energetic incident photon is required for the head-on 
collision of the $\gamma$-photon with the laser pulse.

\begin{figure}
\includegraphics[width=4.0cm]{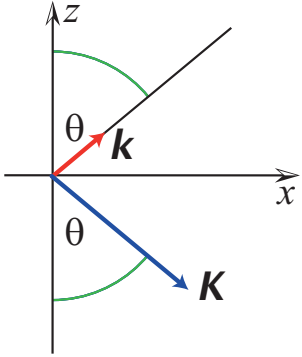}
\caption{(Color online) An arbitrary laboratory geometry (up to the space rotation) for the laser-induced Breit-Wheeler process.}
\label{bwgeom}
\end{figure}

For a fixed energy of the created positron, we find the constraint for its direction:
\begin{equation}
\cos\theta_{\mathrm{e}^+}  > \frac{\sqrt{|\bm{p}_{\mathrm{e}^+}|^2+(m_{\mathrm{e}}c)^2}-K^-}{|\bm{p}_{\mathrm{e}^+}|}.
\label{constrain}
\end{equation}
Hence, it follows that a pair can be created in any space direction if the $\gamma$-photon energy is sufficiently large, i.e.,
\begin{equation}
\omega_{\bm{K}} > \frac{c|\bm{p}_{\mathrm{e}^+}|+\sqrt{(c|\bm{p}_{\mathrm{e}^+}|)^2+(m_{\mathrm{e}}c^2)^2}}{1-\bm{n}\cdot\bm{n}_{\bm{K}}}.
\end{equation}
On the other hand, the pair creation will not occur at all if the incident photon energy $\omega_{\bm K}$ is too small,
\begin{equation}
 \omega_{\bm K}<\frac{-c|\bm{p}_{\mathrm{e}^+}|+\sqrt{(c|\bm{p}_{\mathrm{e}^+}|)^2+(m_{\mathrm{e}}c^2)^2}}{1-\bm{n}\cdot\bm{n}_{\bm{K}}}.
\end{equation}
In the intermediate range of $\omega_{\bm K}$, pairs will be created in a finite region of space.

Let us note that similar constraints can be derived in relation to the electron, if $w$ is redefined such that $w=K-p_{{\rm e}^-}$.

\subsection{Relativistic invariance}
\label{invariance}

The formulas derived above for the differential probability distributions of the laser-induced Breit-Wheeler 
process [Eqs. \eqref{cp10} or \eqref{bwtrain}] are not relativistically invariant. We have 
presented them in a noninvariant form in order to analyze below the validity of an
approximation of a laser pulse by a modulated plane wave.
Since quantum electrodynamics is the relativistic quantum theory, 
therefore, it is appropriate to use the relativistic invariance in order to simplify the geometry 
of the Breit-Wheeler process. In particular, we can show that from the theoretical point of view 
the head-on collision of the $\gamma$-photon with the laser pulse is the generic configuration 
for this process, in the sense that there exists the Lorentz transformation that transforms 
an arbitrary laboratory geometry to the head-on configuration. To this end, let us recall that the pair 
creation is forbidden if the laser beam and the $\gamma$-photon propagate in the same direction (see, Sec.~\ref{kinematics}). 
If so, we can always choose the space coordinates such that the vectors $\bm{k}$ and $\bm{K}$ are 
coplanar with the $(xz)$-plane, with the axes chosen such that ${\bm k}$ and ${\bm K}$ form the same angle 
$\theta$ with the $z$-axis, as presented in Fig. \ref{bwgeom}. We shall assume, without loosing 
the generality, that $0\leqslant\theta <\pi/2$, so that $k_z>0$. Next, let us apply the Lorentz boost, 
$\mathcal{L}_x$, in the direction of the $x$-axis with the parameters $\beta_x$ and $\gamma_x$ such that, 
\begin{equation}
\mathcal{L}_x:\quad \beta_x=\sin\theta, \gamma_x=1/\cos\theta, \label{lorentzx}
\end{equation}
which transforms the laser and the $\gamma$-photon wave four-vectors, $k$ and $K$, to
\begin{align}
k'&=k'_0 (1,0,0,1)=k'_0 n' , \nonumber \\
K'&=K'_0 (1,0,0,-1)=K'_0 n_{\bm{K}'} , \label{trans1}
\end{align}
where $k'_0=k_0\cos\theta$ and $K'_0=K_0\cos\theta$ (we see, that this boost lowers the frequencies of both the laser field 
and the $\gamma$-photon). Also, the created electron and positron four-momenta are transformed accordingly,
\begin{align}
p'_0 &= (p_0-p_x\sin\theta)/\cos\theta , \nonumber \\
p'_x &= (p_x-p_0\sin\theta)/\cos\theta , \nonumber \\
p'_y &= p_y, \quad p'_z=p_z .
\end{align}
Moreover, the real polarization vectors for the laser field, $\bm{\varepsilon}_1$ and $\bm{\varepsilon}_2$, 
as well as the polarization vector of the $\gamma$-photon, $\bm{\varepsilon}_{\bm{K}\sigma}$, are transformed 
appropriately. However, if we choose them as being either parallel or perpendicular to the $(xz)$-plane, then 
they preserve these properties under the Lorentz boost \eqref{lorentzx}.

We can proceed even further and apply the Lorentz boost in the $z$-direction, $\mathcal{L}_z$, with the parameters:
\begin{equation}
\mathcal{L}_z:\quad \beta_z=\frac{k^{\prime 2}_0-k^{\prime\prime 2}_0}{k^{\prime 2}_0+k^{\prime\prime 2}_0}, 
\gamma_z=\frac{k^{\prime 2}_0+k^{\prime\prime 2}_0}{2k'_0k^{\prime\prime}_0}, \label{lorentzz}
\end{equation}
and an arbitrary $k^{\prime\prime 2}_0>0$. One can check that the laser four-vector $k'$ and the $\gamma$-photon four-vector $K'$ transform as
\begin{align}
k^{\prime\prime}&=k^{\prime\prime}_0 (1,0,0,1)=k^{\prime\prime}_0 n' , \nonumber \\
K^{\prime\prime}&=K^{\prime\prime}_0 (1,0,0,-1)=K^{\prime\prime}_0 n_{\bm{K}'} , \label{trans2}
\end{align}
where
\begin{equation}
K^{\prime\prime}_0=\frac{k'_0}{k^{\prime\prime}_0}K'_0.
\end{equation}
This means that such a boost keeps the product of both frequencies invariant; by increasing the laser field 
frequency we decrease the $\gamma$-photon energy, and vice-versa. In other words, the laser field 
cannot be described as low-frequency and the $\gamma$-photons cannot be thought of as the high-energy radiation, 
as compared to the electron rest energy, since these concepts depend on the choice of the reference frame.

For the head-on geometry, we can define the analogue of the energy-angular differential probability distribution 
given by Eq.~\eqref{cp10}. For this geometry and for the Lorentz boost \eqref{lorentzz}, the linear polarization vectors 
$\bm{\varepsilon}'_1$, $\bm{\varepsilon}'_2$, and $\bm{\varepsilon}_{\bm{K}'\sigma}$ do not change and the total 
pair creation probability, $\mathsf{P}^{(\mathrm{p})}(k'\cdot K')$, depends only on the product of $k'$ and $K'$ 
(note that all the scalar products of the wave vectors with the polarization vectors vanish for this configuration). 
Hence, as the relativistically invariant quantity, $\mathsf{P}^{(\mathrm{p})}(k'\cdot K')$ fulfills the equation
\begin{equation}
\mathsf{P}^{(\mathrm{p})}(k'\cdot K')=
\mathsf{P}^{(\mathrm{p})}(k^{\prime\prime}\cdot K^{\prime\prime}).
\end{equation}
Equivalently, instead of $k'\cdot K'$ one can use the $s$-invariant, $s=(k'+K')^2=2k'\cdot K'$, as discussed in Ref.~\cite{Titov2012}.

Now, the total probability of pair creation can be expressed in one of the equivalent forms,
\begin{align}
& \mathsf{P}^{(\mathrm{p})}(k'\cdot K') \nonumber \\
 & = \int \mathrm{d}E_{\bm{p}'_{\mathrm{e}^-}}\mathrm{d}^2\Omega_{\bm{p}'_{\mathrm{e}^-}} \frac{|\bm{p}'_{\mathrm{e}^-}|}{c^2}  \mathsf{D}^{(\mathrm{p})}_{\mathrm{e}^-}(k',K',p'_{\mathrm{e}^-})
\label{electrondistr} \\
 & = \int \mathrm{d}E_{\bm{p}'_{\mathrm{e}^+}}\mathrm{d}^2\Omega_{\bm{p}'_{\mathrm{e}^+}} \frac{|\bm{p}'_{\mathrm{e}^+}|}{c^2}  \mathsf{D}^{(\mathrm{p})}_{\mathrm{e}^+}(k',K',p'_{\mathrm{e}^+}),
\label{positrondistr}
\end{align}
where
\begin{eqnarray}
\mathsf{D}^{(\mathrm{p})}_{\mathrm{e}^-}(k',K',p'_{\mathrm{e}^-})&=&\frac{\alpha (m_{\mathrm{e}}c^2)^2}{(2\pi)^2 k'\cdot p'_{\mathrm{e}^+}}\mathsf{D}^{(\mathrm{p})},\\
\mathsf{D}^{(\mathrm{p})}_{\mathrm{e}^+}(k',K',p'_{\mathrm{e}^+})&=&\frac{\alpha (m_{\mathrm{e}}c^2)^2}{(2\pi)^2 k'\cdot p'_{\mathrm{e}^-}}\mathsf{D}^{(\mathrm{p})},
\end{eqnarray}
and
\begin{equation}
\mathsf{D}^{(\mathrm{p})}=\frac{{k'}^0}{\omega_{\bm{K}'}} \sum_{\lambda_{\mathrm{e}^-},\lambda_{\mathrm{e}^+}=\pm} \Big|\sum_N D_N\frac{1-\mathrm{e}^{-2\pi\mathrm{i}{P'}_N^0/{k'}^0}}{{P'}_N^0} \Big|^2 .
\end{equation}
The reason for writing Eqs.~\eqref{electrondistr} and~\eqref{positrondistr} in their current forms is that the integration measure
\begin{equation}
\mathrm{d}E_{\bm{p}}\mathrm{d}^2\Omega_{\bm{p}} \frac{|\bm{p}|}{c^2}= \frac{\mathrm{d}^3p}{E_{\bm{p}}}
\end{equation}
is relativistically invariant. Hence, the energy-angular distributions for electrons, $\mathsf{D}^{(\mathrm{p})}_{\mathrm{e}^-}$,
and positrons, $\mathsf{D}^{(\mathrm{p})}_{\mathrm{e}^+}$, are also relativistic invariant with respect to the Lorentz 
boost~\eqref{lorentzz}. This invariance, together with the gauge invariance discussed above, has been used for testing 
our numerical code. 

In closing this Section, let us stress that the introduction of the second Lorentz boost was motivated mainly by two reasons. First, in the real experimental 
setup the laser field frequencies are much smaller than the electron rest energy. This might cause some numerical problems. To avoid them,
it is better to analyze numerically this process in the reference frame where the frequency 
of the laser field is of the order of the relativistic unit of energy or it is comparable to the 
frequency of the $\gamma$-photon. Second, the relativistic invariance of the Breit-Wheeler 
process shows that, irrespectively of the value of the laser field frequency in the laboratory 
frame, it is forbidden to consider the laser field as a slowly changing function on the electron's 
Compton wavelength scale; in particular, it is not permitted to approximate the action of the laser 
field by a constant or by a slowly-varying time-dependent electric fields.

\section{Single pulse vs. train of pulses}
\label{compare}

Consider a single laser pulse described by the following four-vector potential,
\begin{equation}
 A(k\cdot x)=A_0B \bigl[\varepsilon_1g_1(k\cdot x)\cos\delta +\varepsilon_2g_2(k\cdot x)\sin\delta\bigr],
\label{a1}
\end{equation}
which propagates along the $z$-axis and lasts for time $T_{\rm p}$. Its fundamental 
frequency equals $\omega=2\pi/T_{\rm p}$ and the wave four-vector is
$k=k^0(1,{\bm e}_z)=(\omega/c)(1,{\bm e}_z)$. We choose two linear polarization vectors, 
$\varepsilon_1=(0,{\bm e}_x)$ and $\varepsilon_2=(0,{\bm e}_y)$. In Eq.~\eqref{a1},
we keep the parameter $\delta$ which describes the ellipticity of the laser field. 
The pulse shape functions $g_1(k\cdot x)$ and $g_2(k\cdot x)$ are chosen such that
\begin{eqnarray}
 g_1'(k\cdot x)&=&-N_f\sin^2\Bigl(\frac{k\cdot x}{2}\Bigr)\sin(N_{\rm osc}k\cdot x+\chi)\nonumber\\
 &=&-N_f\sin^2\Bigl(\frac{k_{\rm L}\cdot x}{2N_{\rm osc}}\Bigr)\sin(k_{\rm L}\cdot x+\chi),\label{a222}\\
 g_2'(k\cdot x)&=&N_f\sin^2\Bigl(\frac{k\cdot x}{2}\Bigr)\cos(N_{\rm osc}k\cdot x+\chi)\nonumber\\
 &=&N_f\sin^2\Bigl(\frac{k_{\rm L}\cdot x}{2N_{\rm osc}}\Bigr)\cos(k_{\rm L}\cdot x+\chi),
\label{a2}
\end{eqnarray}
for $0\leqslant k_{\rm L}\cdot x\leqslant 2\pi N_{\rm osc}$, and they are 0 otherwise. Here,
$\chi$ is the carrier-envelope phase, whereas $k_{\rm L}=N_{\rm osc}k=(\omega_{\rm L}/c)(1,{\bm e}_z)$ with the 
carrier frequency of the laser pulse, $\omega_{\rm L}$. The normalization constant $N_f$
in the above equations is chosen such that~\cite{TwoColor}
\begin{equation}
\frac{1}{2\pi}\int_0^{2\pi}{\rm d}\phi \Bigl([g_1'(\phi)]^2\cos^2\delta+[g_2'(\phi)]^2\sin^2\delta\Bigr)=\frac{1}{2}.
\label{a3}
\end{equation}
The pair creation process induced by pulses of different duration but carrying out the same 
energy will be compared. This condition is satisfied if the peak value of the vector potential~\eqref{a1}
is scaled as $\sqrt{N_{\rm osc}}$ when changing the number of laser field oscillations within
the pulse, $N_{\rm osc}$. If we keep,
\begin{equation}
 \mu=\frac{|eA_0|}{m_{\rm e}c},
\label{a4}
\end{equation}
then $B=\sqrt{N_{\rm osc}}$ has to be chosen in Eq.~\eqref{a1}. Note that Eq.~\eqref{a1}
is a particular realization of a more general vector potential~\eqref{LaserVectorPotential},
with $f_1(k\cdot x)=B g_1(k\cdot x)\cos\delta$ and $f_2(k\cdot x)=B g_2(k\cdot x)\sin\delta$.

Up till now, we have specified a single laser pulse containing the fixed energy but with a varied duration. Next, let us define a train 
of such pulses as their infinite sequence [Eqs.~\eqref{a1},~\eqref{a222}, and~\eqref{a2}], with the zero harmonic subtracted
in Eqs.~\eqref{a222} and~\eqref{a2}. This does not change the value of the electric and the magnetic fields
characterizing the pulse. At the same time, the energy within one pulse from the train is fixed, 
irrespectively of its duration.

For a numerical illustration, we consider a head-on collision of the laser pulse
introduced above with a nonlaser photon. As argued in Sec.~\ref{invariance}, any
kinematic configuration of a laser and a nonlaser fields can be transformed, by
a suitable Lorentz transformation, to a reference frame in which they become 
counterpropagating. This excludes the case when both fields propagate in the same
direction, which however is irrelevant to the present problem (see, Sec.~\ref{kinematics}). For the chosen 
head-on configuration, we take the laser field parameters such that the carrier frequency 
$\omega_{\rm L}$ is $0.1m_{\rm e}c^2$ and $\mu=1$. The carrier-envelope phase of the laser 
field is $\chi=0$ and the parameter $\delta=0$, unless otherwise stated. The incident nonlaser photon is linearly polarized, with
the polarization direction specified below, and the energy $\omega_{\bm K}=2m_{\rm e}c^2$.
As explained in Sec.~\ref{invariance}, due to relativistic invariance, 
the presented results are valid in an arbitrary reference frame transformed according to~\eqref{lorentzz}, 
if the following condition for $\omega_{\rm L}$ and $\omega_{\bm K}$ 
is kept: $\omega_{\rm L}\omega_{\bm K}=0.2 (m_{\rm e}c^2)^2$, and if the electron and the positron momenta are 
changed accordingly. Since $s=(k+K)^2= 4\omega_{\rm L}\omega_{\bm K}/c^2< 2(m_{\mathrm{e}}c)^2$, 
hence, we consider the subthreshold pair creation, according to the nomenclature introduced in \cite{Titov2012}.

\begin{figure}
\includegraphics[width=8.5cm]{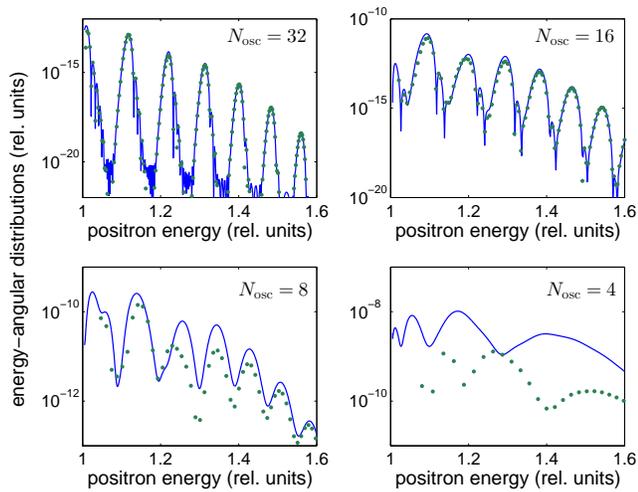}
\caption{(Color online) Comparison of the positron energy-angular distributions 
for a single laser pulse (blue solid line) and for a train of pulses (green dots)
[Eqs. \eqref{bwpulse} and \eqref{bwtrain}, respectively], in the case when a different number 
of laser field oscillations, $N_{\mathrm{osc}}$, is contained in the pulse. The positron polar and azimuthal 
angles are $\theta_{\mathrm{e}^+}=\pi/2$ and $\varphi_{\mathrm{e}^+}=0$, respectively. The results are for the parameters
of the laser pulse such that $\mu=1$ and $\omega_{\rm L}=0.1m_{\rm e}c^2$, and for the
energy of the $\gamma$-photon $\omega_{\bm K}=2m_{\rm e}c^2$. Both, the $\gamma$-photon and the laser field 
 are linearly polarized, with ${\bm\varepsilon}_{\bm{K}\sigma}={\bm \varepsilon}_1=\bm{e}_x$.}
\label{bw2012.07.17}
\end{figure}

\begin{figure}
\includegraphics[width=8.5cm]{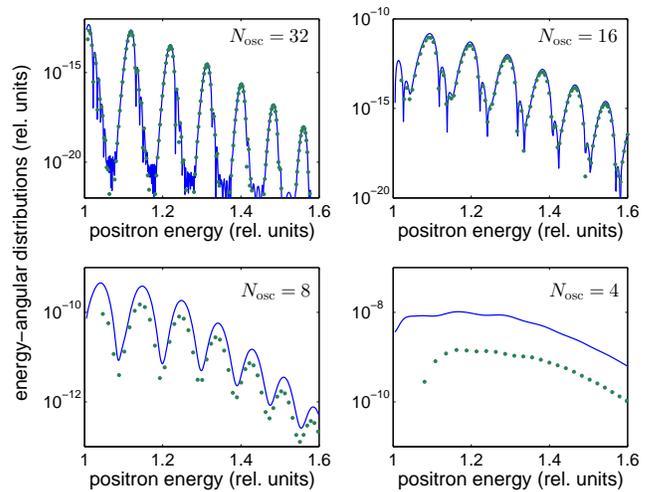}
\caption{(Color online) The same as in Fig. \ref{bw2012.07.17}, but the linear polarization vectors
of the colliding $\gamma$-photon and the laser field are perpendicular to each other, i.e., 
${\bm \varepsilon}_{{\bm K}\sigma}=-{\bm e}_y$ and ${\bm\varepsilon}_1={\bm e}_x$. }
\label{bw2012.07.17p}
\end{figure}

In Fig.~\ref{bw2012.07.17}, we demonstrate the energy distributions for positrons created in the direction
of the laser field polarization ($\theta_{{\rm e}^+}=\pi/2$ and $\varphi_{{\rm e}^+}=0$), 
in the case when the linear polarization vector of the colliding $\gamma$-photon is in the same 
direction as that of the laser field, ${\bm \varepsilon}_{{\bm K}\sigma}={\bm \varepsilon}_1={\bm e}_x$. The solid blue line 
corresponds to the Breit-Wheeler process by a single laser pulse, whereas the green dots are for the Breit-Wheeler
process by a train of pulses [let us remind that in the latter case, the respective distribution is defined
per pulse from a train; for more details, see Secs.~\ref{single} and~\ref{train}, respectively]. The 
spectra in Fig.~\ref{bw2012.07.17} relate to different pulse durations, characterized by a different number of field oscillations, 
$N_{\rm osc}$, as specified in each panel. While for long pulses (including 32 and 16 oscillations 
of the laser field) a very good agreement between the results for an individual pulse and 
a pulse from a train of pulses is observed, this agreement decreases with decreasing the pulse
duration. The same was observed for the energy spectra of Compton photons~\cite{Compton}, showing 
the necessity to account for an actual temporal structure of ultrashort laser pulses when studying
processes in strong laser fields. Moreover, our results show that for long laser pulses it is allowed 
to investigate the process numerically by treating a single laser pulse as a pulse from the train of pulses, 
which significantly speeds up the computations.

The aforementioned feature can also be noted in Fig.~\ref{bw2012.07.17p}. This figure illustrates  
the exact same physical situation as Fig.~\ref{bw2012.07.17}, but for a different linear polarization of the $\gamma$ 
photon, when ${\bm \varepsilon}_{{\bm K}\sigma}=-{\bm e}_y$. It is particularly well seen for $N_{\rm osc}=4$,
that the results in Figs.~\ref{bw2012.07.17} and~\ref{bw2012.07.17p} for an individual laser pulse and a pulse from a sequence 
of pulses do not only differ in magnitude but they are also shifted with respect to each other.
This is caused by a different energy threshold in both of
these cases, that is related to a different laser-field dressing, as defined by Eqs.~\eqref{cp1} and~\eqref{ct4}. 
For long laser pulses [when $\langle f_i\rangle\rightarrow 0$ ($i$=1,2) in Eq.~\eqref{cp1}], the energy threshold 
is basically the same for both the Breit-Wheeler process induced by a single laser pulse and by a laser pulse train. 
With decreasing the pulse duration, the threshold energy starts to differ for both cases which 
results in a respective shift of the energy spectra of created particles.

Let us note that the probability distributions of created positrons, shown in Figs.~\ref{bw2012.07.17} and~\ref{bw2012.07.17p}, 
increase in magnitude when decreasing the pulse duration; for the parameters considered in Figs.~\ref{bw2012.07.17} 
and~\ref{bw2012.07.17p}, one observes roughly four orders of magnitude increase when the driving laser pulse changes 
from 32 to four cycles. Once we keep the energy contained in a laser pulse fixed, the maximum value of the electric 
and magnetic fields must be effectively increased when decreasing the pulse duration. Consequently, a significant 
enhancement of the probability of pair production induced by shorter laser pulses is observed.

\begin{figure}
\includegraphics[width=8.5cm]{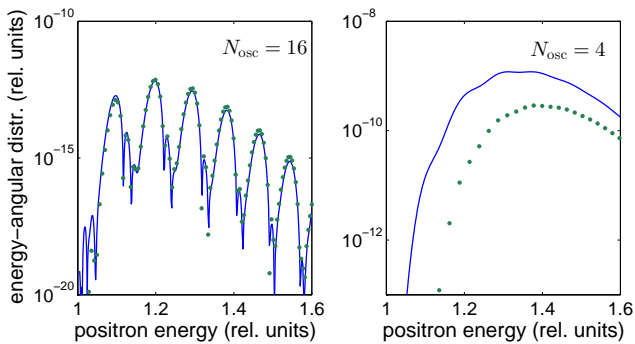}
\caption{(Color online) The same as in Fig. \ref{bw2012.07.17}, but the laser field is circularly 
polarized ($\delta=\pi/4$) whereas the $\gamma$-photon is linearly polarized along the $x$-axis, i.e.,
${\bm \varepsilon}_{{\bm K}\sigma}={\bm e}_x$.}
\label{bwc2012.08.12}
\end{figure}

Although our main objective in this paper is to present the results for the linearly polarized driving
field, in Fig.~\ref{bwc2012.08.12} we show the angular spectra of positrons created by a circularly
polarized laser field [$\delta=\pi/4$ in Eq.~\eqref{a1}] (the remaining parameters are the same as in Fig.~\ref{bw2012.07.17}).
The colliding $\gamma$-photon is still linearly polarized, with the polarization vector ${\bm\varepsilon}_{{\bm K}\sigma}={\bm e}_x$.
We have also checked that for ${\bm\varepsilon}_{{\bm K}\sigma}=-{\bm e}_y$, the results hardly change.
As expected, when comparing Fig.~\ref{bw2012.07.17} and~\ref{bwc2012.08.12}, the probabilities of pair 
creation are smaller for the circularly polarized laser field than for the linearly polarized field. 
Similar strong dependence of pair production on the polarization of a driving laser field 
was observed in Ref.~\cite{KK} for the nonlinear Bethe-Heitler process. 

\begin{figure}
\includegraphics[width=8.5cm]{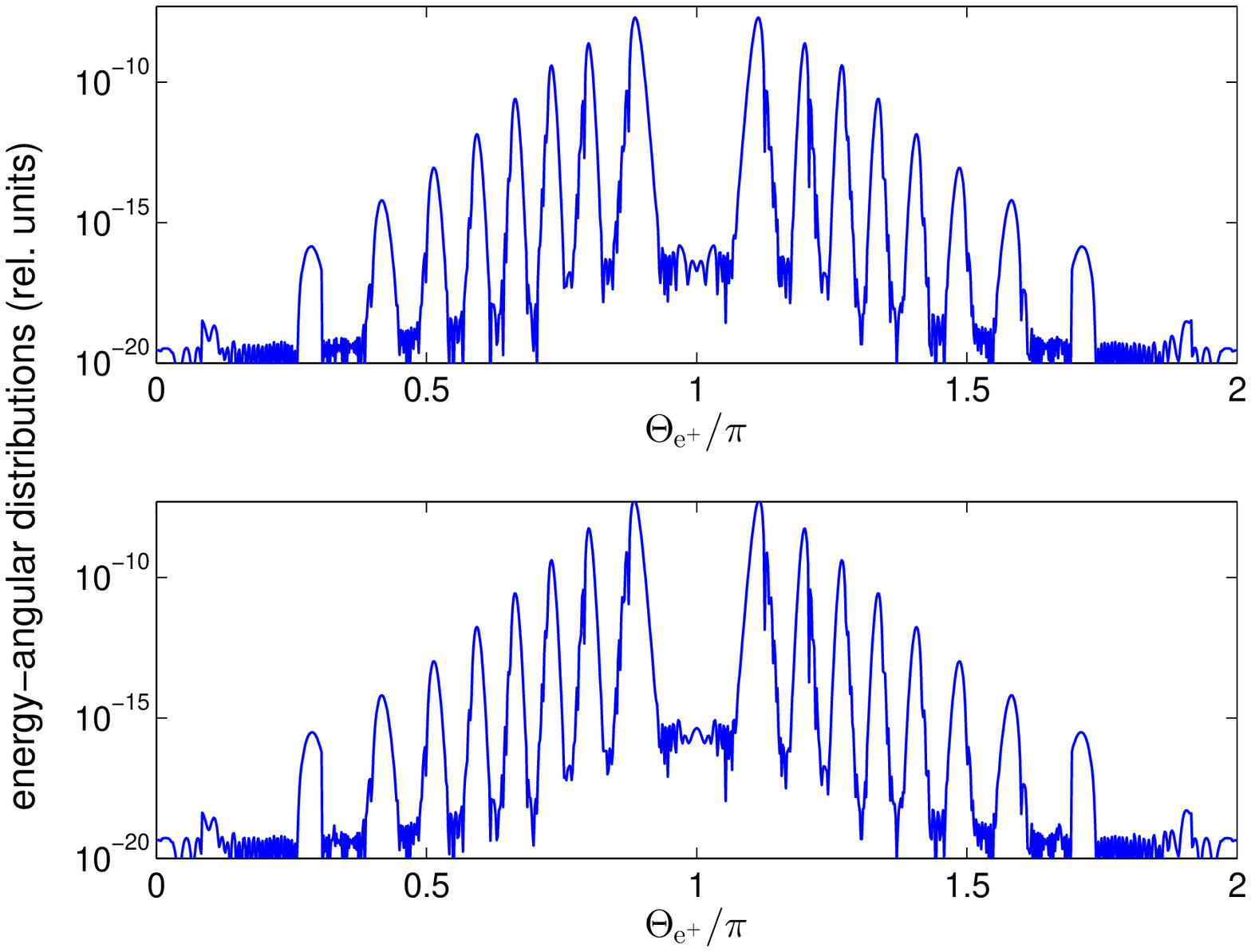}
\caption{(Color online) The angular distributions of positrons produced with momentum $|\bm{p}_{\mathrm{e}^+}|=m_{\mathrm{e}}c$
by a single laser pulse including 32 field oscillations ($N_{\mathrm{osc}}=32$). 
The other parameters are $\mu=1$, $\omega_{\mathrm{L}}=0.1m_{\mathrm{e}}c^2$, $\omega_{\bm{K}}=2m_{\mathrm{e}}c^2$, 
$\varphi_{\mathrm{e}^+}=0$ for $\Theta_{{\rm e}^+}\leqslant \pi$ (or, $\varphi_{\mathrm{e}^+}=\pi$ for 
$\Theta_{{\rm e}^+}>\pi$), and $\chi=0$. The upper panel relates to the case when the $\gamma$-photon is 
linearly polarized such that ${\bm \varepsilon}_{\bm{K}\sigma}=\bm{e}_x$, whereas the lower panel is for 
${\bm \varepsilon}_{\bm{K}\sigma}=-\bm{e}_y$. The peaks appear when $N_{\mathrm{eff}}$ changes by multiples of $N_{\rm osc}$. 
The spectra are symmetric with respect to the propagation direction of the incident $\gamma$-photon.}
\label{bwang32}
\end{figure}

In Fig.~\ref{bwang32}, we present angular distributions for positrons created with momenta $|{\bm p}_{{\rm e}^+}|=m_{\rm e}c$
in the $(xz)$-plane, i.e., for $\varphi_{{\rm e}^+}=0$ or $\pi$, 
in the case when a 32-cycle pulse collides with a countermoving $\gamma$-photon. The nonlaser photon is also linearly polarized, with the
polarization vector either parallel (upper panel) or perpendicular (lower panel) to the $(xz)$-plane.
Note that the positron angle $\Theta_{{\rm e}^+}$, featured in Fig.~\ref{bwang32}, is defined such that 
it is equal to $\theta_{{\rm e}^+}$ when $\varphi_{{\rm e}^+}=0$, or to $2\pi-\theta_{{\rm e}^+}$ 
when $\varphi_{{\rm e}^+}=\pi$. One can see that for long driving laser pulses, the positron angular distributions 
stay roughly the same regardless of the respective polarizations of the pulse and the $\gamma$-photon. 
Namely, they are symmetric with respect to the propagation direction of the 
incident $\gamma$-photon, showing a regular peak structure. These peaks appear when $N_{\rm eff}$
changes by 32, i.e., by the number of laser field oscillations within the pulse. In other words, 
the observed spectra are typical for a multiphoton process with absorption of an integer number of $\omega_{\rm L}$-photons.
While the highest peaks are observed for angles $\Theta_{{\rm e}^+}$ close to $\pi$, they gradually decrease in magnitude when
$\Theta_{{\rm e}^+}$ changes towards $0$ or $2\pi$. In fact, there is hardly any pair production in the direction
of the pulse propagation, in this particular reference frame. Let us also note that the angular distributions for accompanying 
electrons look similar to Fig.~\ref{bwang32}, which is true only for long driving laser pulses.

\section{Carrier-envelope phase effects}
\label{CEP}

In this Section, we investigate the effect of the carrier-envelope phase of a few-cycle pulse on angular spectra
of created positrons in the nonlinear Breit-Wheeler process. This is done for different configurations
of linear polarizations of the colliding laser pulse (i.e., with $\delta=0$) and the nonlaser photon.

\begin{figure}
\includegraphics[width=8.5cm]{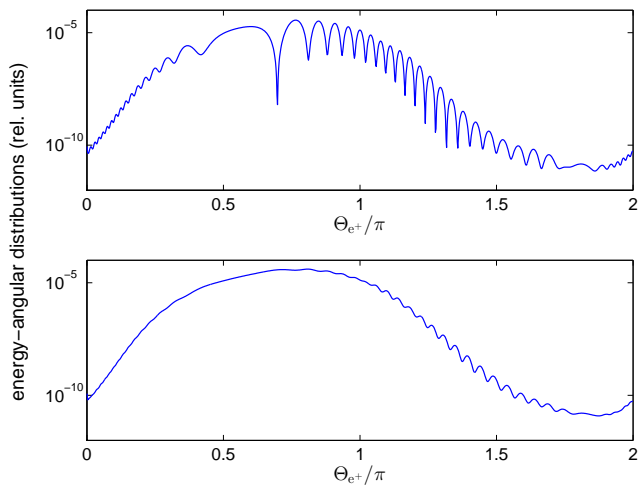}
\caption{(Color online) The same as in Fig. \ref{bwang32} but for $N_{\mathrm{osc}}=2$. 
The angular spectra are asymmetric with respect to the propagation direction of the colliding $\gamma$-photon.}
\label{bwang2f0}
\end{figure}

\begin{figure}
\includegraphics[width=8.5cm]{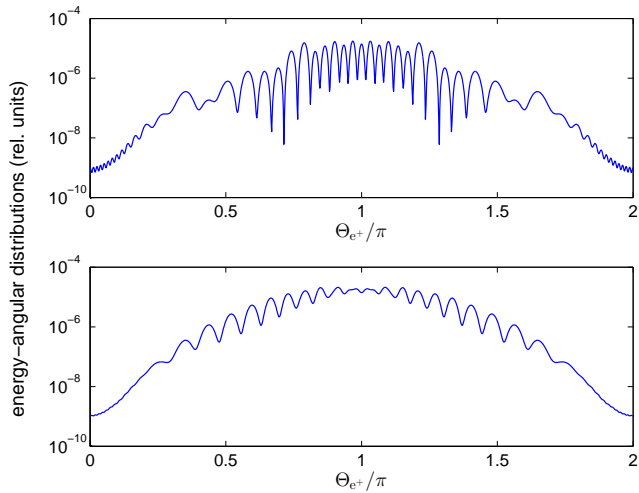}
\caption{(Color online) The same as in Fig. \ref{bwang2f0} but for a carrier-envelope phase of the
driving laser pulse $\chi=\pi/2$. This time, the angular distributions are symmetric with 
respect to the propagation direction of the $\gamma$-photon.}
\label{bwang2f5}
\end{figure}

In Figs.~\ref{bwang2f0} and~\ref{bwang2f5}, we demonstrate the angular distributions of positrons with momenta
$|{\bm p}_{{\rm e}^+}|=m_{\rm e}c$ created in a head-on collision of a two-cycle pulse ($N_{\rm osc}=2$) with a $\gamma$-photon.
As before, the spectra are for the positron azimuthal angle, $\varphi_{{\rm e}^+}=0$ for $0\leqslant\Theta_{\mathrm{e}^+}<\pi$ 
and $\varphi_{{\rm e}^+}=\pi$ otherwise. The upper 
panels in both figures relate to the collinear configuration of the linearly polarized
vectors ${\bm\varepsilon}_{{\bm K}\sigma}={\bm\varepsilon}_1={\bm e}_x$, whereas the lower panels are 
for the perpendicular configuration, when ${\bm\varepsilon}_{{\bm K}\sigma}=-{\bm e}_y$ and ${\bm\varepsilon}_1={\bm e}_x$.
While the results presented in Fig.~\ref{bwang2f0} are for the carrier-envelope phase $\chi=0$,
the results in Fig.~\ref{bwang2f5} correspond to $\chi=\pi/2$. For such choices of $\chi$,
the CEP effect is clearly visible. 


One sees from Fig.~\ref{bwang2f0} that for $\chi=0$, positron angular distributions are asymmetric with respect 
to the propagation direction of the $\gamma$-photon, with its maximum shifted towards angles
$\Theta_{{\rm e}^+}$ smaller than $\pi$. On the other hand, the positron angular spectra 
are symmetric for $\chi=\pi/2$, as demonstrated in Fig.~\ref{bwang2f5}. These features are 
independent of the respective polarization directions of both the laser pulse and the nonlaser photon.
What does depend on their polarization configuration is the oscillatory pattern shown by the
positron angular distributions. In particular, for the collinear polarizations the angular 
spectra of positrons exhibit stronger interferences than for the perpendicular polarizations.
However, the envelope of these oscillatory patterns stays roughly the same. Such a difference in the 
oscillatory behaviors can be intuitively understood by noticing that the positron energy-angular distributions 
depend, in principle, on the positron momentum $\bm{p}_{\mathrm{e}^+}$ through the 
following scalar products: $\bm{k}\cdot \bm{p}_{\mathrm{e}^+}$, $\bm{K}\cdot \bm{p}_{\mathrm{e}^+}$, 
$\bm{p}_{\mathrm{e}^-}\cdot \bm{p}_{\mathrm{e}^+}$, $\bm{\varepsilon}_1\cdot \bm{p}_{\mathrm{e}^+}$ and 
$\bm{\varepsilon}_{\bm{K}\sigma}\cdot \bm{p}_{\mathrm{e}^+}$. For the $\gamma$-photon polarization vector perpendicular 
to the $(xz)$-plane (bottom panels of Figs.~\ref{bwang2f0} and~\ref{bwang2f5}) the last scalar product vanishes. 
Hence, such a behavior has a purely geometrical origin, as our numerical investigations for other configurations show.

In closing this Section, let us compare Figs.~\ref{bwang32} and~\ref{bwang2f0}, which are for the 
same CEP ($\chi=0$) but for different incident pulse durations (32- and two-cycle laser pulses, 
respectively). One can see that the angular distributions of created
positrons are very sensitive not only to a change of the carrier-envelope phase, which is particularly
important for ultrashort laser pulses, but also to a change of the driving pulse duration. For the fixed
CEP, the positron angular spectra are symmetric for very long laser pulses driving the pair 
creation (Fig.~\ref{bwang32}), whereas they become asymmetric with decreasing the pulse duration 
(except the cases when $\chi=\pi/2$ or $3\pi/2$, as illustrated in Fig.~\ref{bwang2f5}). 
The same was observed in Ref.~\cite{Compton} when analyzing the angular spectra of Compton photons
by finite laser pulses. It can be considered, therefore, a general feature of QED processes induced 
by laser pulses of finite durations. The reason being that while for many-cycle laser pulses
the vector potential is symmetric, meaning that $\langle f_i\rangle\sim 0$ for $i=1, 2$
[see, Eq.~\eqref{LaserVectorPotential}], thus for few-cycle laser pulses it is asymmetric (for almost all values of $\chi$). 
In addition, we observe that by decreasing the pulse duration the angular spectra of
positrons presented in Figs.~\ref{bwang32} and~\ref{bwang2f0} exhibit weaker interferences.

\section{Conclusions}
\label{conclusions}

In the present article, the laser-induced Breit-Wheeler scenario of electron-positron pair production 
in which the pairs are created in collisions of a laser field with a nonlaser photon has been investigated. 
The driving laser field has been modeled as a finite laser pulse, along the lines developed in Ref.~\cite{Compton} for
the Compton scattering. In order to check the validity of our theory, a comparison with a modulated
plane wave approximation for the laser field has been performed. Let us stress that we have developed
a general formulation of the laser-induced Breit-Wheeler process, which accounts for arbitrary parameters 
of both the colliding laser pulse (e.g., arbitrary pulse shapes, pulse durations, strengths, etc.) and the nonlaser photon 
(e.g., arbitrary polarization and energy).

Dependence of energy-angular distributions of created particles on parameters of the driving laser pulse
has been studied in this paper in great detail. Contrary to the most recent work by Titov {\it et al.}~\cite{Titov2012}, 
we have focused here on the nonperturbative regime of pair creation. In this regime, a sensitivity of the process 
to the polarization of the colliding laser pulse has been demonstrated, with the highest 
probabilities in the case of linear polarization. Therefore, our main focus in this paper has been on analyzing the case 
when a linearly polarized laser pulse collides with a $\gamma$-photon. Moreover, for numerical illustrations,
we have chosen the head-on configuration of the colliding beams. As we have also shown, any other configuration 
can be obtained by a suitable Lorentz transformation from the head-on setup, and vice-versa.

We have demonstrated that for sufficiently long laser pulses driving the Breit-Wheeler process,
the energy-angular spectra of created particles agree very well with the spectra calculated
for an infinite train of laser pulses, as we refer to using the modulated plane wave approximation.
This agreement breaks down, however, for short pulse durations (see, also Ref.~\cite{Compton}
for the related analysis of Compton scattering). For short pulses, a careful treatment
of their actual temporal structure must be carried out. To demonstrate this, we have studied 
the carrier-envelope phase effects in the signal of pair creation. We have observed that by
changing the CEP, the angular distributions of positrons (electrons) vary dramatically from being symmetric to
asymmetric. This general feature does not depend on the relative linear polarizations
of the colliding laser pulse and the nonlaser photon. Indeed, only a very detailed pattern of positrons (electrons)
spectra depend on their relative space configuration. We believe that the sensitivity of the energy-angular spectra of created particles to a variation
of the CEP can be used as means of phase control. In this context, it would be further interesting to analyze 
the CEP effects in total probabilities of pair creation, or in the electron-positron correlations
that have been defined for the nonlinear Bethe-Heitler process in Ref.~\cite{KK1}. These tasks, 
when performed in the nonperturbative regime, require very demanding numerical calculations.
For this reason, we will perform them separately from this paper and present their results in due course.

\section*{Acknowledgments}

This work is supported by the Polish National Science Center (NCN) under Grant No. 2011/01/B/ST2/00381. 
K.K. gratefully acknowledges the hospitality of the Department of Physics and Astronomy at the University of 
Nebraska, Lincoln, USA, where part of this article was prepared.

\end{document}